\documentclass[preprint2]{aastex}

\shorttitle{Multiscale Dynamics of Solar Magnetic Structures}
\shortauthors{Uritsky et al.}

\begin{document}

\title{Multiscale Dynamics of Solar Magnetic Structures}

\author{Vadim M. Uritsky}
\affil{CUA at NASA Goddard Space Flight Center, Greenbelt, MD 20771 USA}
\email{vadim.uritsky@nasa.gov}

\author{Joseph M. Davila}
\affil{NASA Goddard Space Flight Center, Greenbelt, MD 20771 USA}

\begin{abstract}
Multiscale topological complexity of solar magnetic field is among the primary factors controlling energy release in the corona, including associated processes in the photospheric and chromospheric boundaries. We present a new approach for analyzing multiscale behavior of the photospheric magnetic flux underlying this dynamics as depicted by a sequence of high-resolution solar magnetograms. The approach involves two basic processing steps: (1) identification of timing and location of magnetic flux origin and demise events (as defined by \citet{deforest07}) by tracking spatiotemporal evolution of unipolar and bipolar photospheric regions, and (2) analysis of collective behavior of the detected magnetic events using a generalized version of Grassberger - Procaccia correlation integral algorithm. The scale-free nature of the developed algorithms makes it possible to characterize the dynamics of the photospheric network across a wide range of distances and relaxation times. Three types of photospheric conditions are considered to test the method: a quiet photosphere, a solar active region (NOAA 10365) in a quiescent non-flaring state, and the same active region during a period of M-class flares. The results obtained show (1) the presence of a topologically complex asymmetrically fragmented magnetic network in the quiet photosphere driven by meso- and supergranulation, (2) the formation of non-potential magnetic structures with complex polarity separation lines inside the active region, and (3) statistical signatures of canceling bipolar magnetic structures coinciding with flaring activity in the active region. Each of these effects can represent an unstable magnetic configuration acting as an energy source for coronal dissipation and heating. 

\end{abstract}


\section{Introduction}

The solar magnetic field is often seen as the underlying cause of solar activity and the primary agent coupling the Sun to the heliosphere. High-resolution observations show that it possesses an incredible level of complexity across all relevant spatial and temporal scales (e.g., \citet{schrijver92, balke93, lawrence93, cadavid94, abramenko02, abramenko10, georgoulis05, mcateer10, parnell09}). Understanding the origin and the physical significance of this complexity is among the most fundamental unresolved problems of contemporary solar science. Its further analysis requires accurate quantification of cross-scale plasma processes involved in the formation and evolution of the observed multiscale magnetic patterns.

Solar magnetic complexity is deeply rooted in the life cycle of coronal perturbations. Coronal magnetic field is known to accumulate progressively large amounts of magnetic energy and helicity before producing explosive phenomena such as flares and coronal mass ejections (see e.g. \citet{solanki06}). The emergence of complex domains of magnetic connectivity accompanying this process is instrumental as it makes the magnetic free energy available for dissipation \citep{antiochos98, schrij05, edmondson10}. The initial build-up of the magnetic complexity can be driven by the injection of new magnetic structures (e.g., \citet{falconer08, abramenko10, mcateer10, conlon10}), or by the fragmentation of the existing flux subject to turbulent footpoint shifting both in solar active regions (see e.g. \citet{abramenko05a}) and in the quiet Sun (\citet{lamb08, chaouche11} and refs therein). 

The resulting magnetic patterns featuring mixed-polarities regions with strongly sheared fields are prone to kinetic instabilities which can unfreeze the magnetic flux and start magnetic reconnection. The reconnection brings the system to a lower energy state and reduces the magnetic complexity locally but usually not globally \citep{aschwanden06}. Cooperative interactions between multiple reconnecting plasma tubes may lead to the second generation of multiscale phenomena at the coronal level, including self-organized criticality \citep{lu91}, storms of nanoflares in individual multistranded coronal loops \citep{klimchuk06, viall11, morales08}, and cascades of instabilities involving multiple reconnecting loops \citep{hughes03, torok11}. The distortions in the topology of open field regions caused by some of these effects can be a major factor controlling both slowly varying and transient solar wind \citep{antiochos07}. In the quiet Sun, the wide ranges of spatial and temporal scales manifested by the brightness of the soft X-ray and EUV corona are likely to reflect the multiscale geometry and dynamics of the underlying photospheric network \citep{falconer98}. The importance of solar magnetic complexity in triggering coronal dissipation events and the possibility of their self-organization into multiscale structures is confirmed by many theoretical models (see, for instance, \citet{charbonneau01, aschwanden02, aschwanden11}). 

Gaining a deeper insight into the nature of solar magnetic complexity and its impact on coronal instabilities requires innovative data analysis tools focused on cause and effect links between two or more interconnected multiscale processes. There is a serious need to explore their cross-correlation properties, in addition to the autocorrelation properties studied before. Furthermore, the causality of solar activity is an inherently spatiotemporal phenomenon which unfolds both in spatial and temporal domains of analysis. Until now, the spatiotemporal aspect of solar magnetic complexity has not been given proper attention in observational studies. Thus, for example, the algorithms of solar magnetic tracking commonly used for analyzing the emergence, cancellation \citep{martin84}, fragmentation and coalescence dynamics of the photospheric network are essentially monoscale. The outputs of these algorithms and the interpretation of the obtained results depend, explicitly or implicitly, on {\it ad hoc} choices of spatial and temporal correlation scales (see \citet{deforest07} for a detailed review). To our knowledge, none of the methods proposed so far can simultaneously address short- and long-range properties of solar magnetic network within the same data analysis framework while treating both spatial and temporal scales as free parameters. The available post-processing techniques, such as Fourier spectrograms, spatial correlation functions, or multifractal statistics,  share the same methodological "blind spot" -- they cannot adequately describe transient cross-scale coupling associated with time-dependent nature of studied multiscale patterns.

In this paper, we propose a new data analysis paradigm focused on multiscale spatiotemporal properties of the photospheric flux dynamics. As opposed to previously proposed methods for identifying correlated magnetic events, the developed algorithms contain no {\it a priori} assumptions regarding the intrinsic spatial and temporal interaction scales. By accepting this strategy, we take into account the fact that magnetic concentration regions, acting as nodes of the photospheric network, can communicate over arbitrary long distances beyond the local neighborhood. We also assume that these  communications can encompass a wide range of response times. 

Our technique involves two main processing steps. First, we determine timings and locations of origin and demise magnetic events \citep{deforest07} based of a direct spatiotemporal tracking of unipolar and bipolar photospheric regions above a specified magnetic flux threshold. Second, we analyze pairwise correlations between the detected magnetic events using a generalized Grassberger - Procaccia correlation integral technique. By design of the algorithms, any detected characteristic scale is an attribute of the studied data and not that of the method. 

This proposed data analysis framework is successfully tested on a sequence of high cadence SOHO MDI magnetograms representing a solar active region (NOAA 10365) and a quiet Sun. The application of spatiotemporal methodology extends earlier statistical analyses of multiscale geometry of the photospheric field (e.g., \cite{lawrence93, mcateer10}), and allows us to evaluate scaling characteristics of the dynamics of the photospheric flux not reported before. The results obtained reveal significant difference in the network dynamics underlying flaring and non-flaring regimes of NOAA 10365, and deliver a detailed picture of spatiotemporal clustering in the adjacent quiet photosphere. 
 
The paper has the following structure. Section 2 provides the description of the studied data, solar conditions, and applied statistical tools. Section 3 presents main results of our analysis and their qualitative interpretation in terms key multiscale plasma processes operating in solar photosphere and corona. Section 4 summarizes our findings and highlights some remaining problems.

\section{Data and methods}

\subsection{Observational data and solar conditions}

\begin{figure*}[htbp]
\begin{center}
\noindent\includegraphics*[width=12 cm]{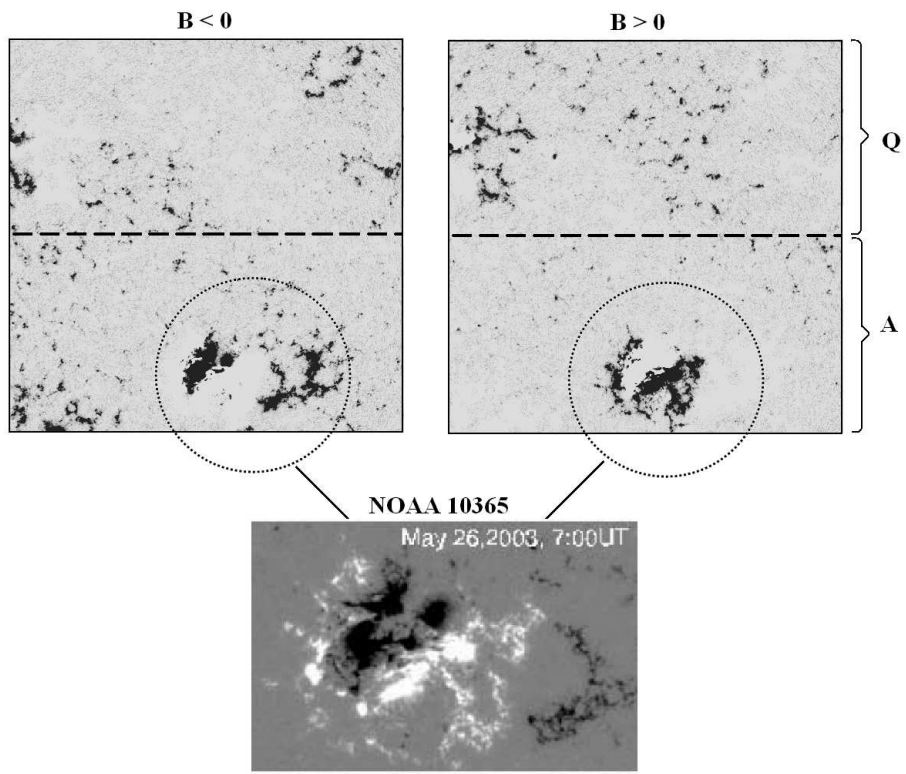} 
\caption{\label{fig3} SOHO MDI magnetogram (14:00 05/26/2003) showing the studied field of view (approx. 445$\times$445 Mm, centered at the solar disk center) including the subregions with quiet (active) conditions above (below) the dashed horizontal line representing solar equator. The NOAA 10365 active region is outlined by the dotted circular line.}
\end{center}
\end{figure*}

We analyzed a sequence of 816 high-resolution SOHO MDI magnetograms \citep{scherrer95} collected between 14:00 05/26/2003 and 04:35 05/27/2003, average cadence time 64 seconds. The images were calibrated using standard SolarSoft procedures and rebinned down to the 0.9$\times$0.9 Mm resolution to reduce the pixel noise, with the final image size 512$\times$512 pixels. The studied field of view contains two adjacent spatial domains representing different levels of solar activity (Fig. \ref{fig3}).

The quiet region to the north of the solar equator (labeled by letter Q in Fig. \ref{fig3}) exhibits a characteristic pattern of strongly fragmented magnetic flux loosely organized into an irregular cellular structure typical of a quiet Sun. As we discuss below, this structure is associated with photospheric meso- and supergranulation, and it demonstrates a significantly geometric asymmetry of positive and negative magnetic polarities precluding direct identification of canceling bipolar pairs. The region operates in a statistically steady state as evidenced by the similarity of scaling properties of origin and demise dynamics across the studied range of scales. 

The southern portion of the studied field of view (labeled with A in the figure) is dominated by the NOAA 10365 active region which first appeared on the visible solar disk on 19 May 2003 at heliographic latitude $\sim -5^{\circ}$. A few days later, a new bipolar structure formed inside the active region. By merging with the decayed preexisting flux, the new dipole developed a complex polarity separation line \citep{verbeeck11}. Starting from 25 May, the total magnetic flux of NOAA 10365 was increasing steadily before reaching a plateau on 29 May 2003. The region gradually rotated counterclockwise with the leading negative flux polarity \citep{tian11}. The continuous injection of a new flux in the northern portion of NOAA 10356 created a $\delta-$spot configuration with a positive region progressively surrounded by the emergent negative flux \citep{chandra09}. The magnetic field pattern of the emerging bipole featured two large elongated areas of opposite polarity suggesting a twisted flux tube of positive helicity \citep{tian11, chae04}. 

The activity of the NOAA 10365 region underwent significant changes during the studied time interval. Between 15:00 05/26/2003 and 20:48 05/26/2003, the active region generated multiple M-class flares, with the GOES 12 1.0-8.0 {\AA} flux remaining above the $10^{-10}$ watts m$^{-2}$ level. During 20:48 05/26/2003 - 02:42 05/27/2003, the GOES flux stayed consistently below this level indicating no noticeable flare activity. The two intervals were analyzed separately in order to compare magnetic clustering effects in flaring and non-flaring regimes.

\subsection{Definition of magnetic events}

To determine timings and positions of magnetic events in the studied set of magnetograms, we invoked direct spatiotemporal tracking approach \citep{uritsky07, uritsky10} in which magnetic features living for more than one image sampling interval were treated as subvolumes of the three-dimensional space-time. The history of each feature is represented by a connected set of image pixels containing an unsigned magnetic field above a specified detection threshold $B_{th}$. 

The idea of the threshold-based spatiotemporal tracking has been implemented in the context of solar magnetograms by \citet{berger96, berger98}. The object tracking method used in their study was based on measuring the centroid locations of the most compact and intense magnetic features. Here we invoke a somewhat different approach which does not rely on the centroid location as a guiding center for feature tracking. Time-evolving magnetic concentrations are treated as true 3-dimensional objects, which makes it possible to study a more general class of magnetic structures whose positions cannot be adequately described by a centroid.

In order to overcome the inherent memory limitations of standard cluster analysis algorithms such as, e.g., the Label Region function of the Interactive Data Language (IDL), we have developed a new numerical scheme enabling fast decomposition of a multidimensional data array into a set of connected pixels (clusters) while dramatically reducing the amount of stored information. 

The first step of our technique consists of building a table of so-called "activations" -- contiguous time intervals during which the condition $ |B| > B_{th}$ is fulfilled in each of the image pixels. 
The second step is to find and label spatially connected clusters of the activations. This is done using the "breadth-first search" principle which avoids backtracking of search trees representing individual clusters. We find that it is important to consider all 8 nearest neighbors of each spatiotemporal node, including the diagonal neighbors, when identifying connected activations. Finally, the table of activations is rearranged according to the assigned cluster labels, in order to provide faster access to the detected structures. The output data array contains the complete information on the location, shape, and time evolution of all the detected magnetic concentration regions in the studied 3D volume.

It can be seen that the proposed algorithmic solution reverses the usual order of feature tracking steps (feature segmentation within in an individual image followed by cross-frame feature identification within an image set, see e.g. \citet{deforest07}). Instead, we start with detecting temporal traces of features in individual photospheric locations, and then consider their spatial adjacency. We found this algorithmic solution to be remarkably fast and memory-efficient, with the size of the output data array being typically by two order of magnitudes smaller compared to the original data. The proposed scheme is obviously easy to parallelize. A more complete description of the developed algorithm and its numerical validation on high-Reynolds number magnetohydrodynamic turbulence can be found in \citep{uritsky10a}.

Using the described tracking technique, we identified spatial and temporal coordinates of origin and demise magnetic events described by positive and negative line-of-sight magnetic polarity. The origin events (OE) represent newly detected magnetic concentration regions which just exceeded the threshold. The demise events (DE) occur when a previously found region becomes undetectable, i.e when its flux content drops below $B_{th}$. This classification is analogous to the subdivision into "origin" and "demise" events described by \cite{deforest07} although these authors used a different feature tracking method. It should be emphasized that our definition of events is not limited to the injection of new magnetic bipoles into the photosphere and their subsequent submergence into the convection zone. In essence, we deal with an apparent dynamics associated with crossings of the detection threshold $B_{th}$ by local magnetic field. While the resulting sets of events may contain a fraction of truly emerging and submerging magnetic bipoles, they also represents magnetic flux coalescence, fragmentation, cancellation, and other effects leading to spatiotemporal variability of the photospheric field.

The photospheric velocities of positive an negative magnetic spots inside the NOAA 10365 active region were in the range 0.07 to 0.24 km/s \citep{chandra09}, or less than 0.015 Mm per minute. For the time scales $\tau < 20$ min studied here, this motion would produce a spatial displacement of $\leq 0.2$ Mm which is below the spatial resolution of SOHO MDI data. Based on this estimate, the magnetic flux configuration of the active region can be considered quasi-static in the photospheric plane, with most of the observed changes owing to enhancement and reduction of local photospheric magnetic flux rather than its horizontal convective motion. The same conclusion applies to the quiet region. Typically, the quiet Sun  horizontal speeds lie in the range 0.45 - 0.50 km/s \citep{shine00}; over a 20-minute interval, this motion is likely to be limited to a single MDI pixel and remain unresolved.

To reduce the level of granulation noise \citep{deforest07}, we focused on the magnetic features which lasted more than 5 minutes and encompassed a sufficiently large spatial area ($>$ 4 Mm$^2$). The valid features were also required to have small initial and final linear sizes ($\leq$ 3 Mm) ensuring accurate determination of their locations. The magnetic features which did not meet these quality constraints as well as those truncated by the boundaries of the chosen field of view or  observing interval were excluded from further analysis.


\subsection{The generalized correlation integral (GCI) analysis}

The structure of multiscale data sets such as the ones generated by solar magnetograms can be accurately described using the family of order-$q$ generalized fractal dimensions \citep{grassberger83a}:
\begin{eqnarray}\label{eq0}
D_q &=& \left\{
\begin{array}{rl}
\displaystyle{ 
\frac{\partial \mbox{ log } \sum_{i=1}^{n(\epsilon)}{p_i \mbox{log}\, p_i}   }{\partial \mbox{ log } \epsilon} , q=1 
}\\
\displaystyle{ 
\frac{1}{q-1} \frac{\partial \mbox{ log } \sum_{i=1}^{n(\epsilon)}{p_i^q}   }{\partial \mbox{ log } \epsilon} , \mbox{otherwise}. 
}\\
\end{array} \right.
\end{eqnarray} 
Here, $n(\epsilon)$ is the number of non-empty hyper-cubes of a partitioning with length scale $\epsilon$, and $p_i$ is the probability for a data point to fall into a cube $i$ of the partitioning; see \citet{schuster05} for a pedagogical introduction to the concept.  Among the dimensions defined by eq.(\ref{eq0}), the Hausdorff fractal dimension ($q=0$), the information dimension ($q=1$), and the correlation dimension ($q=2$) are widely used. We base our analysis on the latter because it represents a second-order statistical moment which can be naturally converted into a cross-correlation measure of bipolar magnetic structures as shown below.

For a set of point objects embedded in a multidimensional metric space, the correlation dimension is commonly evaluated using the Grassberger - Procaccia correlation integral \citep{grassberger83}. The correlation integral (CI) is computed by counting the number $C$ of pairs of set elements separated by the Euclidean distance $<\ell$. The shape of the $C(\ell)$ dependence characterizes  clustering effects in the studied set. For a random set with statistically independent element locations $C(\ell) \propto l^{\:d}$, where $d$ is the embedding dimensions. For example, if the elements are uniformly scattered on a smooth surface ($d=2$), the probability to find a neighbor within a given distance $\ell$ scales as the area of a circle with this radius, and hence $C \propto \ell^{\:2}$. If the elements are clustered there exists an interval of scales in which $C(\ell) \propto \ell^{\:\alpha < d}$. 
The power-law exponent $\alpha$ of the correlation integral is an empirical proxy for the correlation dimension. If $\alpha$ is constant across a substantial range of scales (one ore more orders or magnitude), the studied set demonstrates self-similar or self-affine density variations involving voids and aggregates of various sizes. The discrepancy between the correlation dimension $\alpha$ and the embedding dimension $d$ is a sensitive measure of this clustering behavior. 

The CI technique can be generalized to the case when the studied set contains a mixture of elements of two distinct types \citep{kantz94}. In our study, the two types of events are associated with positive and negative line-of-sight polarity magnetic events (origin or demise) embedded in space and in time. The events are described by a combination of spatial coordinates ${\bf r_i}$ and timings $t_i$, $i=1,..,N$, where $N = N_{+} + N_{-}$ is the total number of positive and negative events. The subsets associated with OE and DE dynamics are treated independently. 

As the preparatory step, we compute the matrix ${\bf L} = \{L_{ij}\}$ of pairwise distances $L_{ij} \equiv |{\bf r}_i-{\bf r}_j|$ 
and the vector ${\bf T} = \{T_{ij}\}$ of pairwise time intervals $T_{ij} \equiv |t_i-t_j|$ between all the events, where $i, j \in [1, N]$. 
We also define the binary vector ${\bf P}=\{P_{i}\}$ taking the value 1 (0) if the $i$th event is of positive (negative) type. 
Next, we construct the inclusion operator
\begin{eqnarray}\label{eq1}
{\bf \Psi}_{ij}(\ell,\tau) \equiv \theta(\ell - L_{ij}) \theta(\tau - T_{ij})
\end{eqnarray}
in which $\theta$ is the Heaviside step function. Note that ${\bf \Psi}_{ij}(\ell,\tau)$ equals 1 if the two events $i$ an $j$ lie within spatial distance $\ell$ and their timing difference does not exceed $\tau$, and takes the value 0 otherwise.
Finally, we define the following set of generalized correlation integrals (GCIs):
\begin{eqnarray}\label{eq3}
C_{+}(\ell,\tau)   \!\!\!\!\!\! & = & \!\!\!\!\!\!   {\frac{1}{(N_{+})^2}} \sum_{i,j} P_i P_j {\bf \Psi}_{ij}(\ell,\tau) \label{eq3.1} \\
C_{-}(\ell,\tau)   \!\!\!\!\!\! & = & \!\!\!\!\!\!   {\frac{1}{(N_{-})^2}} \sum_{i,j} (1\!\!-\!\!P_i)(1\!\!-\!\!P_j) {\bf \Psi}_{ij}(\ell,\tau) \label{eq3.2} \\
C_{\pm}(\ell,\tau)  \!\!\!\!\!\! & = & \!\!\!\!\!\!  {\frac{1}{(N_{+})(N_{-})}} \sum_{i,j} P_i(1\!\!-\!\!P_j) {\bf \Psi}_{ij}(\ell,\tau) \label{eq3.3}
\end{eqnarray}

\begin{figure}[htbp]
\begin{center}
\noindent\includegraphics*[width=7.5 cm]{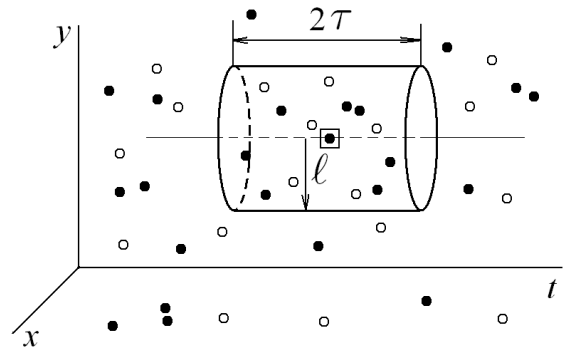} 
\caption{\label{fig1} Schematic illustration of the generalized correlation integral (GCI) technique for $d=2$. Black (white) circles show positive (negative) events embedded in the $d+1$-dimensional space-time. The cylinder represents a spatiotemporal neighborhood centered around the event outlined by the square and defined by a particular choice of $\ell$ and $\tau$. The outlined (positive) event has $N_{+}=7$ unipolar and $N_{\pm}=6$ bipolar neighbors. The GCI calculation consists of an ensemble-averaging over all such neighborhoods centered successively at each of the events in the studied sets, for all relevant combinations of $\ell$ and $\tau$.}
\end{center}
\end{figure}

Each of the GCI functions ($C_{+}$, $C_{-}$, and $C_{\pm}$) approximates the probability of finding a {\it pair} of events of a particular type (respectively positive, negative, and bipolar) within the spatiotemporal neighborhood $V \equiv v_d\: \tau \propto \ell^{\:d}\:\tau$ in which $v_d$ is the volume of the $d$-dimensional hypersphere (Fig. \ref{fig1}). By construction, the autocorrelation integrals $C_{+}$ and $C_{-}$ characterize the internal structure of positive and negative subsets of events and are analogous to the usual CI \citep{grassberger83}, whereas the {\it cross-correlation integral} $C_{\pm}$ describes interactions between positive and negative events in bipolar magnetic pairs. The mathematical definition of $C_{\pm}$ is similar to the cross-correlation sum proposed by \citet{kantz94} for evaluating an overlap of two chaotic attractors embedded in a reconstructed phase space of a nonlinear dynamical system. The primary difference is that here we deal with an essentially anisotropic metric space spanned by spatial and temporal coordinates of solar magnetic events. 

It is easy to show that randomly distributed events with statistically independent positions and occurrence times yield $C \propto V$, $\forall C \in \{C_+, C_-, C_{\pm} \}$. Any departure from this "`null"' scaling regime indicates spatial and/or temporal correlations between the events. By adopting the  scaling ansatz developed in the theory of nonequilibrium critical phenomena \citep{barabasi95}, we assume that $C(\ell)$ and $C(\tau)$ dependences can be factorized after a proper renormalization resulting in a set of spatial and temporal power-laws: 
\begin{eqnarray}\label{eq4}
C_{+} \propto  \ell^{\:\alpha_{+}} \tau^{\:\beta_{+}}\\
C_{-} \propto  \ell^{\:\alpha_{-}} \tau^{\:\beta_{-}}\\
C_{\pm} \propto  \ell^{\:\alpha_{\pm}} \tau^{\:\beta_{\pm}}
\end{eqnarray}
The spatial autocorrelation dimensions $\alpha_{+}$ and $\alpha_{-}$ characterize geometric clustering of respectively positive and negative events; the events of either type are clustered if the corresponding dimension is below $d$. Temporal autocorrelation dimensions $\beta_{+}$ and $\beta_{-}$ have the same meaning in the time domain, with $\beta < 1$ reflecting clustering effects along the time axis (since the embedding dimension of time is 1). 


As opposed to $\alpha_+$, $\alpha_-$, $\beta_+$ and $\beta_-$ dimensions describing unipolar correlations {\it within} the sets of positive and negative magnetic events, the cross-correlation dimensions $\alpha_{\pm}$ and $\beta_{\pm}$ are intended to quantify mutual coupling {\it between} the two types of events. If negative and positive events are statistically independent in space and in time, $C_{\pm}$ scales as $\propto r^{\:d}\:\tau$ similarly to the autocorrelation integrals $C_+$ and $C_-$, and so $\alpha_{\pm}=d$ and $\beta_{\pm}=1$. For correlated sets of bipolar events, the cross-correlation dimensions take values below or above the embedding dimension depending on the sign of the bipolar correlation (positive or negative, correspondingly). 

\begin{figure}[ht]
\noindent\includegraphics*[width=7.5 cm]{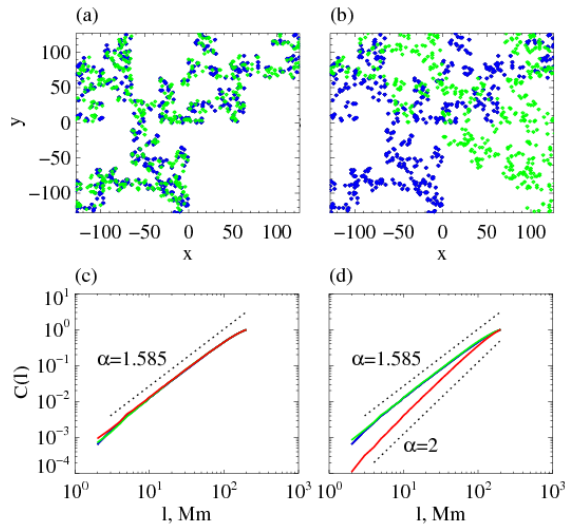} 
\caption{\label{fig2} Identifying auto- and cross-correlations in simulated clustered sets using the GCI approach. The studied sets (blue and green dots) are generated by the same (a) and different (b) realizations of the underlying stochastic Cantor set (shown with gray color) with fractal dimension log(3)/log(2) $\approx$1.158. Bottom panels (c) and (d) show auto- and cross-correlation integrals (eq. \ref{eq3.1} - \ref{eq3.3}) computed for the pairs of sets in (a) and (b), correspondingly. 
The autocorrelation integrals of each set are shown with the matching color (blue or green). The cross-correlation integrals are in red. The code of the stochastic Cantor set is courtesy of G. Uritsky.}
\end{figure}

Asymptotically, the cross-correlation dimensions are independent of the intrinsic clustering structure of the compared sets \citep{kantz94}. In other words, if each of the unipolar sets is clustered but the sets are unrelated to each other, we expect that $\alpha_{\pm} \approx d$ and $\beta_{\pm} \approx 1$ for sufficiently large $N$ and negligible finite-size effects (so that $\ell$ and $\tau$ are much smaller than the largest relevant spatial and temporal scales). Fig. \ref{fig2} illustrates this property of the GCI statistics on static sets of bipolar events constructed by using the stochastic Cantor set algorithm \citep {turcotte89} with the theoretical fractal dimension $d_F =$log(3)/log(2). In panel (a), positive and negative polarity sets are generated by randomly sampling the same underlying Cantor set. They have a statistically similar shape, and all three correlation integrals ($C_+$, $C_-$ and $C_{\pm}$) collapse onto a single curve, with the log-log slope approaching $d_F$ as expected. 
In panel (b), the compared sets are created by sampling two independent runs of the Cantor algorithm. These sets have the same autocorrelation properties as in Fig.\ref{fig2}(a) but must be uncorrelated with each other. Indeed, the measured $\alpha_{+}$ and $\alpha_{-}$ dimensions are close to the theoretical value $d_F$, whereas $\alpha_{\pm}$ approaches the embedding dimension $d=2$ confirming mutual statistical independence of the two sets. The GCI method leads to this correct conclusion despite the complex structure of the compared sets generating false visual similarities at various scales.

For the sake of brevity, the experimental part of this paper focuses on spatial scaling properties of magnetic events as reflected by $\alpha$ exponents computed at different $\tau$ scales. The delay time is varied but treated as a parameter rather than as an independent scaling variable, with the main emphasis placed on the multiscale geometry of unipolar and bipolar magnetic structures as well as the mechanisms of their cross-scale interaction and organization. The analysis of $\beta$ exponents and the associated temporal clustering is left for future research.

To estimate $\alpha$ dimensions, we evaluated log-log slopes of the corresponding GCI functions using  a least-square linear regression model built for several selected ranges of spatial scales. In addition to this approach yielding a discrete set of $\alpha$ values, we used the continuous scalogram technique introduced by \citet{uritsky11} which enables classification of clustering regimes across the entire studied range of scales. The GCI scalograms are defined as 

\begin{equation}\label{eq5} 
\alpha(\ell, \tau) = \frac{\partial \mbox{ log } C(\ell, \tau)   }{\partial \mbox{ log } \ell},
\end{equation}
\noindent
where $\alpha \in \{\alpha_+, \alpha_-, \alpha_{\pm}\}$ and $C \in \{C_+, C_-, C_{\pm}\}$. As in the case with the discrete GCI dimensions, the scalograms (\ref{eq5}) were computed independently for the OE and DE populations of magnetic events.

The statistics reported in the next section are for a single detection threshold $B_{th} = 150$ G. Most of our conclusions holds for a wide range of $B_{th}$ within 1-4 standard deviations above the mean magnetic field magnitude, with the exception of the cross-correlation dimension of the active region exhibiting a more subtle $B_{th}$ dependence as discussed in Section 3.3. 


\section{Results and discussion}

\subsection{Overview of scaling properties of the magnetic events}

Our analysis has shown that the cooperative behavior of the same and mixed polarity magnetic events depends strongly on their spatial and temporal separation as well as on the local solar activity level. It has been also found that bipolar events of either type (origin or demise) exhibit considerably different spatial correlation patterns compared to the corresponding type of the unipolar events. The difference is statistically significant and has a tendency to increase with coronal activity.

\begin{figure*}
\begin{center}
\noindent\includegraphics*[width=11.0 cm]{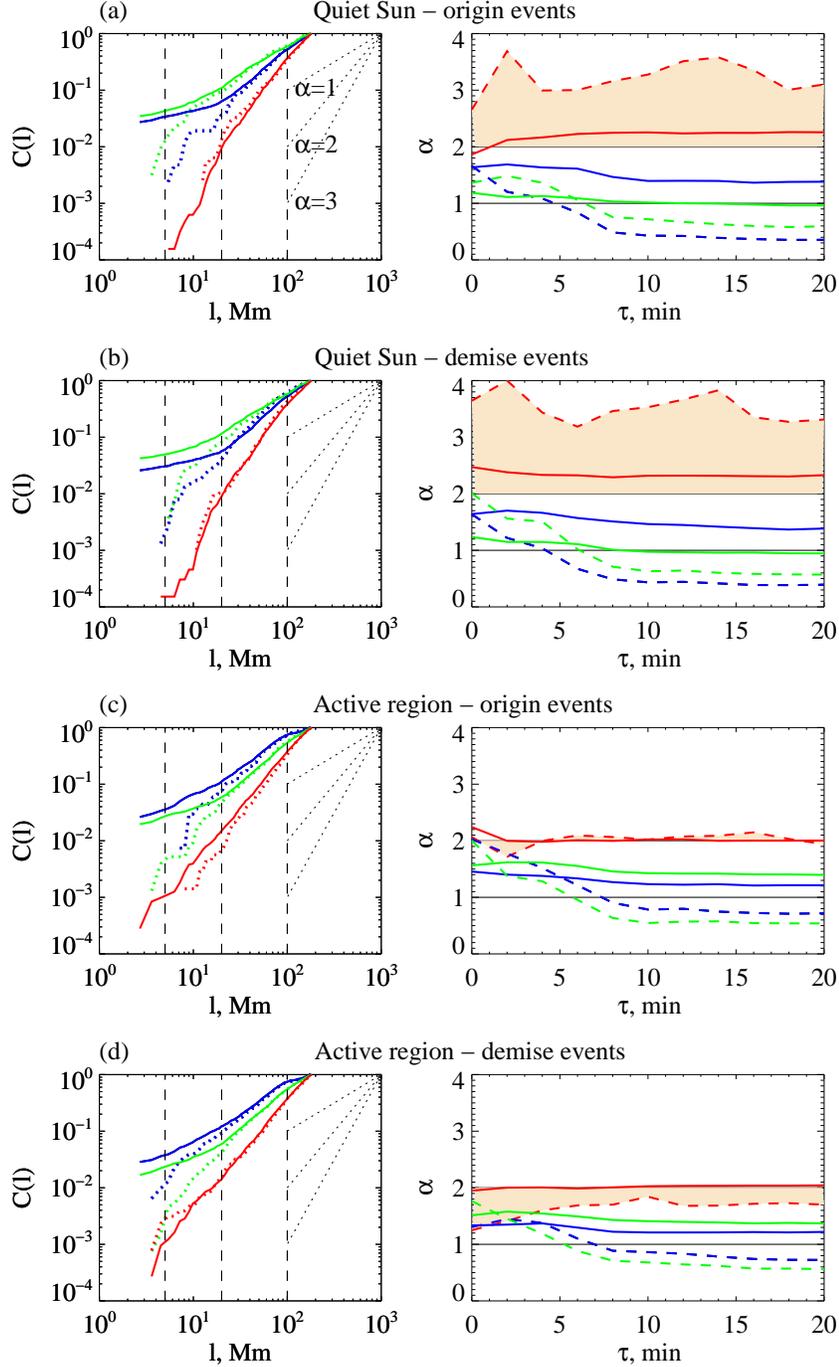} 
\caption{\label{fig4} Results of the GCI analysis of origin and demise activity in quiet and active photospheric regions. Left panels: $C_{+}$, $C_{-}$ and $C_{\pm}$ plots for two time scales $\tau=0$ (10) min shown with dotted (solid) lines. Three integer-valued log-log slopes are shown for reference, with $\alpha=2$ marking uncorrelated regimes. The boundaries of the two scaling ranges of interest associated with the spatial organization above and below the characteristic scale $\ell_0 = 15$ Mm shown with vertical gray lines. Right panels: Dependence  of the dimensions $\alpha_{+}$, $\alpha_{-}$ and $\alpha_{\pm}$ on $\tau$, in two ranges of spatial scales: $\ell<\ell_0$ (solid lines) and  $\ell>\ell_0$ (dashed lines). Color coding: blue (green) -- unipolar $C$ and $\alpha$ statistics of positive (negative) events, red -- bipolar statistics of mixed polarity events. } 
\end{center}
\end{figure*}

Fig. \ref{fig4} presents the combined results of the CGI analysis of OE and DE activity detected in the quiet and active photospheric regions. The left panels show the CGI functions (eq. (\ref{eq3.1}) - (\ref{eq3.3})) evaluated for two fixed time scales, $\tau=0$ min (dotted lines) and $\tau=10 $ min (solid lines). It can be seen that the shape of the autocorrelation integrals characterizing unipolar magnetic clusters changes dramatically with the delay time. For $\tau  = 10$min, both $C_{+}$ and $C_{-}$ exhibit a distinct concave crossover ($\partial ^2 C / \partial {\ell}^2 > 0$) in the vicinity of $\ell=\ell_0\sim 15$ Mm. As we discuss below, this spatial scale can be associated with the lower size limit of photospheric supergranules. 

The right panels in the same figure show the dependence of the GCI dimensions measured below (dashed lines) and above (solid lines) the crossover scale $\ell_0$ on the time scale $\tau$. The systematic decrease of the autocorrelation dimensions $\alpha_{+}$ and $\alpha_{-}$ is likely a result of the $P$-mode contamination due to the five-minute Doppler oscillations leaking into the Zeeman signal and/or the granulation noise described by roughly the same time scale \citep{deforest07}. Each of these mechanisms can modulate the magnetic flux causing the event detection algorithm to loose tracks of weak magnetic features with a field magnitude close $B_{th}$, and to rediscover them after one disturbance cycle. In either case, the increase of $\tau$ beyond the 5-minute time scale leads to better statistical representation of long-living photospheric structures. It is worth noticing that the cross-correlation integrals $C_{\pm}$ (shown in red) are generally steeper than $C_{+}$ and $C_{-}$, and unlike the autocorrelation integrals reveal no systematic dependence on $\tau$. 

Table \ref{table1} summarizes our measurements of the CGI dimensions for the quiet Sun as well as for the active region operating in the quiescent and flaring states. Sections 3.2 and 3.3 below provide a detailed discussion of the scaling properties characterizing these three regimes.

\begin{table*}
\begin{center}
\caption{\label{table1} Estimated values of GCI dimensions of origin (OE) and demise (DE) magnetic events in the quiet (Q) and active (A) regions.}
\begin{tabular}{lcccccc}
\hline
       &         & $\ell < \ell_0$ & & & $\ell > \ell_0$  & \\
Region & $\alpha_{+}$  &  $\alpha_{-}$  & $\alpha_{\pm}$ & $\alpha_{+}$  &  $\alpha_{-}$  & $\alpha_{\pm}$  \\
\\ 
\hline
\\
\multicolumn{7}{c}{{\it OE, $\tau = 0$ min}}\\
Q & 1.66 $\pm$ 0.20& 1.36 $\pm$ 0.06& 2.65 $\pm$ 0.41& 1.64 $\pm$ 0.01& 1.19 $\pm$ 0.02& 1.86 $\pm$ 0.01\\
A & 2.03 $\pm$ 0.27& 2.01 $\pm$ 0.13& 2.06 $\pm$ 0.18& 1.46 $\pm$ 0.01& 1.56 $\pm$ 0.01& 2.24 $\pm$ 0.02\\
\\
\multicolumn{7}{c}{{\it DE, $\tau = 0$ min}}\\
Q & 1.64 $\pm$ 0.13& 2.01 $\pm$ 0.18& 3.64 $\pm$ 0.39& 1.64 $\pm$ 0.01& 1.24 $\pm$ 0.02& 2.47 $\pm$ 0.01\\
A & 1.30 $\pm$ 0.07& 1.77 $\pm$ 0.05& 1.25 $\pm$ 0.06& 1.33 $\pm$ 0.01& 1.51 $\pm$ 0.01& 1.95 $\pm$ 0.02\\
\\
\multicolumn{7}{c}{{\it OE, $\tau = 10$ min}}\\
Q & 0.43 $\pm$ 0.02& 0.72 $\pm$ 0.02& 3.28 $\pm$ 0.13& 1.40 $\pm$ 0.01& 1.01 $\pm$ 0.01& 2.25 $\pm$ 0.01\\
A & 0.78 $\pm$ 0.02& 0.55 $\pm$ 0.02& 2.02 $\pm$ 0.06& 1.23 $\pm$ 0.01& 1.43 $\pm$ 0.00& 2.02 $\pm$ 0.01\\
\\
\multicolumn{7}{c}{{\it DE, $\tau = 10$ min}}\\
Q & 0.43 $\pm$ 0.01& 0.63 $\pm$ 0.02& 3.53 $\pm$ 0.15& 1.46 $\pm$ 0.01& 0.97 $\pm$ 0.00& 2.32 $\pm$ 0.01\\
A & 0.86 $\pm$ 0.01& 0.68 $\pm$ 0.02& 1.84 $\pm$ 0.07& 1.21 $\pm$ 0.01& 1.41 $\pm$ 0.00& 2.03 $\pm$ 0.01\\
\\
\multicolumn{7}{c}{{\it OE, $\tau = 20$ min}}\\
Q & 0.35 $\pm$ 0.02& 0.59 $\pm$ 0.02& 3.10 $\pm$ 0.11& 1.38 $\pm$ 0.01& 0.96 $\pm$ 0.01& 2.26 $\pm$ 0.01\\
A & 0.72 $\pm$ 0.01& 0.54 $\pm$ 0.02& 1.94 $\pm$ 0.06& 1.21 $\pm$ 0.01& 1.40 $\pm$ 0.01& 2.00 $\pm$ 0.00\\
\\
\multicolumn{7}{c}{{\it DE, $\tau = 20$ min}}\\
Q & 0.39 $\pm$ 0.01& 0.57 $\pm$ 0.02& 3.31 $\pm$ 0.12& 1.38 $\pm$ 0.01& 0.95 $\pm$ 0.01& 2.33 $\pm$ 0.01\\
A & 0.72 $\pm$ 0.01& 0.56 $\pm$ 0.02& 1.70 $\pm$ 0.06& 1.22 $\pm$ 0.01& 1.37 $\pm$ 0.00& 2.04 $\pm$ 0.01\\
\hline
\end{tabular}
\end{center}
\end{table*}


\subsection{Quiet Sun}

As can be seen from Fig. \ref{fig4}(a,b) and Table \ref{table1}, unipolar dimensions $\alpha_{+}$ and $\alpha_{-}$ of the quiet photosphere are systematically below the value 2, revealing clustering effects at the levels of meso- and supergranulation. At distances $\ell < \ell_0$, these dimensions are in the range 0 to 1 for both origin and demise events. At larger spatial scales, the unipolar dimensions take values between 1 and 2. The observed scaling crossover uncovers two distinct clustering regimes characterizing the geometry of the photospheric flux below and above the supergranular scale $\ell_0$. The small-scale dimensions are most likely to represent the tendency of the magnetic concentration regions to align along the boundaries of the photospheric supergranules. The non-integer $\alpha$ values indicate the presence of gaps and clusters of various sizes analogous to fractal clustering in a one-dimensional Cantor dust model \citep{turcotte89}. The large-scale clustering at $\ell > \ell_0$ described by $\alpha_-, \alpha_+ \in [1, 2]$ reflects a more isotropic spatial arrangement of solar magnetic flux. At these scales, the flux should be organized into quasi-two-dimensional supergranular structures also containing multiscale voids and clusters.

It is of interest to relate the obtained statistics with commonly discussed theoretical scenarios of photospheric supergranulation. The supergranular flow is known to contain a dominant horizontal component (e.g., \cite{ leighton62, hathaway00}). Its leading role in the formation of the magnetic network outlining supergranular cell boundaries has been intensively discussed \citep{hagenaar97, schrij97, srikanth00}. In deterministic explanatory models, photospheric network appears as a result of forced rearrangement of small-scale surface magnetic fields by a large-scale horizontal flow \citep{martin88, lisle00, wang96}. In stochastic models, the supergranulation network is generated by 
kinematic motion of elementary flux tubes transported passively by the supergranular flow superposed with random displacements produced by granulation \citep{simon95, simon01}, by mutual advection of thermal downflows taking place at random locations \citep{rast03}, or as a self-organization of many interacting magnetic elements following a random walk pattern \citep{crouch07}.

The deterministic models operate with a well-ordered supergranular flow which would result in integer $\alpha $ values both at small and large spatial scales, contrary to our observations. The second (stochastic) group of models seems to be in a qualitative agreement with the correlation analysis results. Our measurements show that the network is significantly fragmented at the supergranulation scale and exhibits an approximate co-alignment of magnetic flux elements with the lines of merging of the adjacent supergranular flow cells. The fractional values of $\alpha_{-}$ and $\alpha_{+}$ exponents at both small and large scales can be a footprint of a stochastic generating process, possibly related to the aggregation, fragmentation, and cancellation of randomly shuffled magnetic concentration regions as proposed by \citet{crouch07}. 

To some extent, our estimates can be affected by the Doppler noise due to the flow motion at the supergranulation scale. In contrast to the short-scale noise sources mentioned in section 3.1, the supergranulation flows are capable of injecting additive signals into MDI magnetograms over a wide range of scales (see e.g. \citet{meunier10} and refs therein). However, we expect this mechanism to have a limited effect on our results. The rms values of the erroneous magnetic field caused by Doppler noise are known to increase with mean magnetic field. The error should be quite small for the magnetic field magnitudes comparable with the chosen detection threshold $B_{th}$. The simulation study by \citet{settele02} predicts SOHO MDI rms errors to be less than 10 G for a relatively strong mean signal of 800 G subject to 5-minute velocity oscillations with amplitude up to $\sim$ 300 m/s. Extrapolating their estimates to $B_{th}$ = 150 G yields an uncertainty of $\sim$ 1 G which can be safely neglected. This conclusion applies to both quiet Sun and AR 10365 active region since we applied the same detection threshold on both sides of the solar equator. 

Furthermore, the supergranulation Doppler noise exhibits strong dependence on the heliocentric angle from disk center. At large heliocentric angles, the line-of-sight motion is dominated by the horizontal component (100-300 m/s, see e.g. estimates by \citet{welsch12} based on local correlation tracking) which can lead to a sizable Doppler effect. At smaller angles, this motion is represented by the much smaller radial velocity component ranging from 7 to 30 m/s \citep{williams11, hathaway02}. Our tests have shown that all main features of the GCI statistics discussed in this section, including the crossover scale $\ell_0$, are clearly present in this central solar region where the supergranulation Doppler noise plays only a minor role. 

\begin{figure}[htbt]
\noindent\includegraphics*[width=7.5 cm]{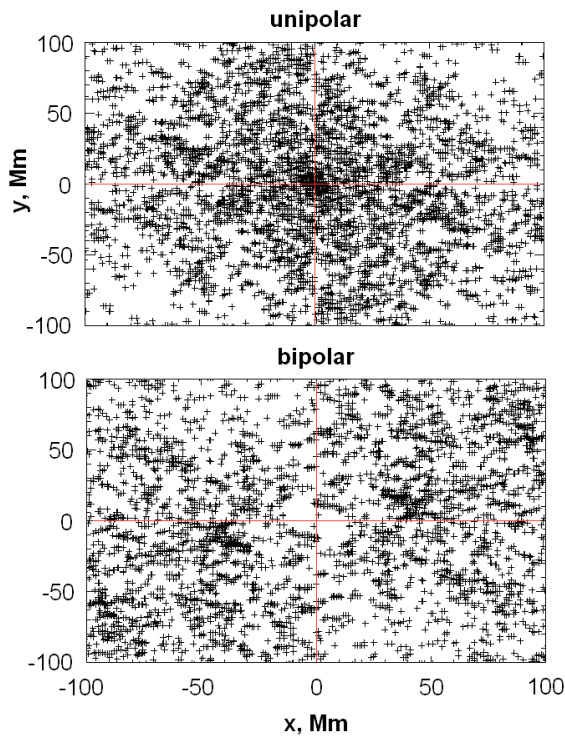} 
\caption{\label{fig5} Unipolar and bipolar clustering maps of the quiet photospheric region, $M_{up}$ and $M_{bp}$, visualizing the difference in the geometry of the OE dynamics of the same (left) and opposite (right) magnetic polarity. } 
\end{figure}

Unlike the autocorrelation (unipolar) dimensions, the $\alpha_{\pm}$ dimensions of the quiet photosphere are systematically greater than 2 pointing at a consistently negative cross-correlation between opposite polarity events. To visualize the clustering geometries underlying the auto- and cross-correlation results, we constructed the unipolar and bipolar clustering maps defined respectively as $M_{up} \equiv \{ {\bf r}_i - {\bf r}_k | P_i = P_k \}$ and $M_{bp} \equiv \{ {\bf r}_i - {\bf r}_k | P_i \neq P_k \}$. In essence, the maps are obtained by adjusting the origin of the coordinate system consequently with each of the $N$ events, and superposing the results in a single scatterplot while keeping track of the polarity of the events. As expected, the two maps are significantly different in shape (Fig. \ref{fig5}). The dense core of the unipolar map represents the tendency of the magnetic events of the same sign to cluster into compact groups described by $\alpha_{\pm} < 2$. The empty central region of the bipolar map reflects negative correlations inside the bipolar pairs leading to $\alpha_{\pm} > 2$, with the probability to find a neighbor of an opposite polarity increasing with the distance up to $\ell \sim 30-40$ Mm.

The observed anti-correlation of the bipolar events in the quiet-Sun network implies a significant imbalance of the apparent evolution of the magnetic flux in this region. Since the actual flux dynamics must be balanced ($\nabla \cdot {\bf B} =0$), our statistics manifest a geometric asymmetry of newly injected or canceling bipoles precluding the detection of a matching pole. It is known that an asymmetric coalescence of a preexisting unresolved unipolar flux into magnetic concentrations that are sufficiently strong to be detected can produce unbalanced OEs: if two or more like-polarity ends of previously unresolved flux tubes come together, the average field strength can overcome the detection threshold for that polarity only \citep{lamb08}. Likewise, asymmetric fragmentation can lead to unbalanced demise events. Such sign-dependent effects are consistent with our analysis of the quiet solar region. We find that the positive events occurring in this region are by a factor of 1.5 - 3.0 (depending on $\tau$ and $B_{th}$) more frequent than negative events indicating that the positive magnetic flux was significantly more fragmented than the negative flux, which could naturally lead to the described artifacts. This interpretation is also supported by the fact that the dimension $\alpha_+$ is systematically lower than $\alpha_-$ (see Table \ref{table1}) suggesting that spatial arrangement of positive polarity events along supergranule boundaries is more sparse compared to that of negative events.

In addition to being a potential source of false event detection, the sign-dependent coalescence - fragmentation dynamics imply that on average, each negative event is magnetically connected with more than one positive event. The higher-order correlations associated with such magnetic structures are beyond the scope of applicability of the proposed second-order statistical measures focusing on pairwise bipolar interactions such as those in magnetic loops. Addressing such correlations in future studies would be a challenging but rather important task. 

It is also noteworthy that in spite of the geometric asymmetry of positive and negative polarity fluxes, the origin and demise GCI dimensions of either magnetic polarity obtained for the quiet Sun are roughly the same. This can be a manifestation of a steady-state balance between the rate of injection of the magnetic flux from the convection zone, and the rate of its dissipation via coronal heating. It is generally accepted that the network magnetic fields play an important role in heating the solar corona. The total X-ray brightness of an active region is closely connected with the overall length of neutral lines and the presence of strong magnetic shears in the neighboring area \citep{falconer97}. Active core fields of the network of the quiet corona can be the main drivers of bulk  coronal heating and dissipation as evidenced by the correlation between the location of bright X-ray emission points and bipolar magnetic structures \citep{falconer98}. Placed in this context, the nearly identical origin and demise GCI statistics strongly suggest that the coronal heating in the quiet Sun can be a magnetically driven process in which the sources and the sinks of the magnetic free energy share the same long-range correlation structure.

The possibility of the evolving but statistically stationary quiet Sun network such as the one revealed by our analysis has been predicted earlier by \citet{simon01} based on kinematic modeling. Their results show that the total unsigned magnetic flux on the solar surface would decay within a few days due to the annihilation of oppositely directed fields brought together by granular and supergranular flows, unless it is continually replenished at an appropriate constant rate. This requirement was presumably fulfilled in the studied quiet solar region as suggested by the statistically identical scaling behavior of the OE and DE dynamics.

\subsection{Active region NOAA 10365}

The correlation integral analysis of the active region leads to a rather different picture in which the origin and demise magnetic flux events follow distinct scaling laws, with $\alpha_{\pm} \approx 2$ for the former and $\alpha_{\pm} < 2$ for the latter events, see Fig. \ref{fig4}(c,d). 

The $\alpha_{\pm} \approx 2$ condition describing the OE population suggests the absence of sizable spatial correlations in newly detected bipolar pairs. This can be a signature of turbulent dynamics in the convection zone or some other process leading to a highly fragmented polarity separation line. Such lines are often associated with a sustained shearing motion of the field leading to a complex mixture of opposite magnetic polarities and resulting in the nonpotentiality of the active region coronal field \citep{schrij05}. Typically, the heating and flaring of coronal loops above an active region requires magnetic polarity inversion near at least one of the loop footpoints, and is greatly enhanced in the presence of a strong magnetic shear in the core magnetic field \citep{falconer97}. The disordered topology of the shear suggested by our CGI measurements may play a vital role in initiating magnetic reconnection above solar active regions \citep{antiochos98, podgorny07,torok11}.

In agreement with the proposed interpretation, strong shearing flows were observed near the main polarity inversion line of NOAA 10365 \citep{chae04}. The region showed a systematic build-up of magnetic complexity en route to a flare-productive unstable configuration. The onset of coronal activity of NOAA 10365 was preceded by a pronounced distortion of the shape of the main bipole separation line  \citep{verbeeck11}, an injection of the helicity in the polarity inversion region \citep{chae04}, and a sharp increase of the Ising energy \citep{ahmed10} of the bipolar magnetic flux \citep{verbeeck11}. A multi-wavelength analysis of NOAA 10365 confirms the formation of a highly twisted magnetic flux tube structure prior to M-class flares in the NOAA 10365 region. Up to 50$\%$ of the helicity injection rate was contributed by short-lived magnetic structures with characteristic times 10-60 minutes \citep{tian11}. The precursive process outlined above could easily generate a rather complex photospheric field pattern with highly entangled magnetic polarities. In fact, the value $\alpha_{\pm} \approx 2$ implies that the complexity of the magnetic network of the active region was close to the theoretical limit in which the emergence locations of oppositely directed flux elements are totally uncorrelated.

\begin{figure*}
\begin{center}
\noindent\includegraphics*[width=12.0 cm]{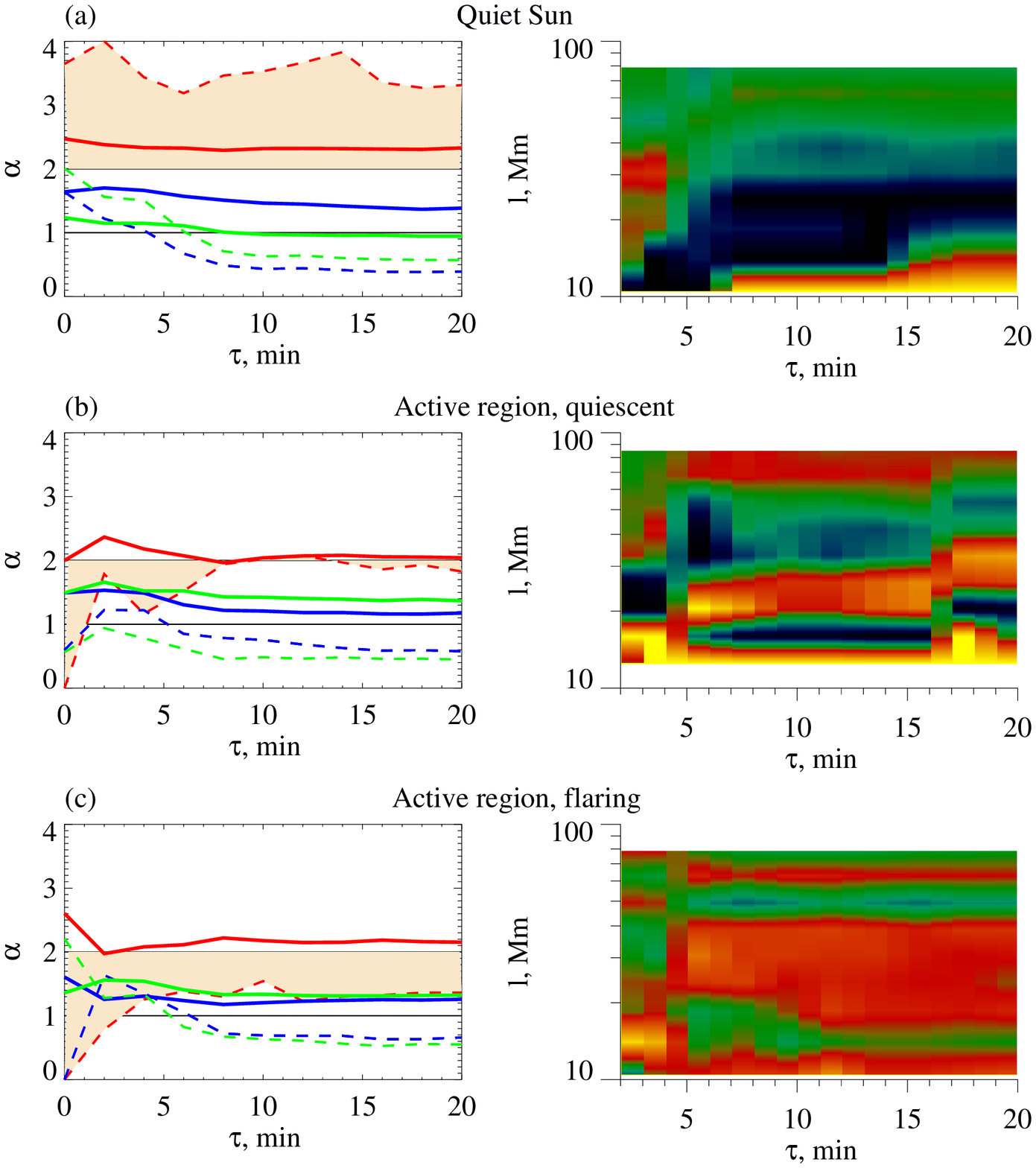} 
\noindent\includegraphics*[width=12.0 cm]{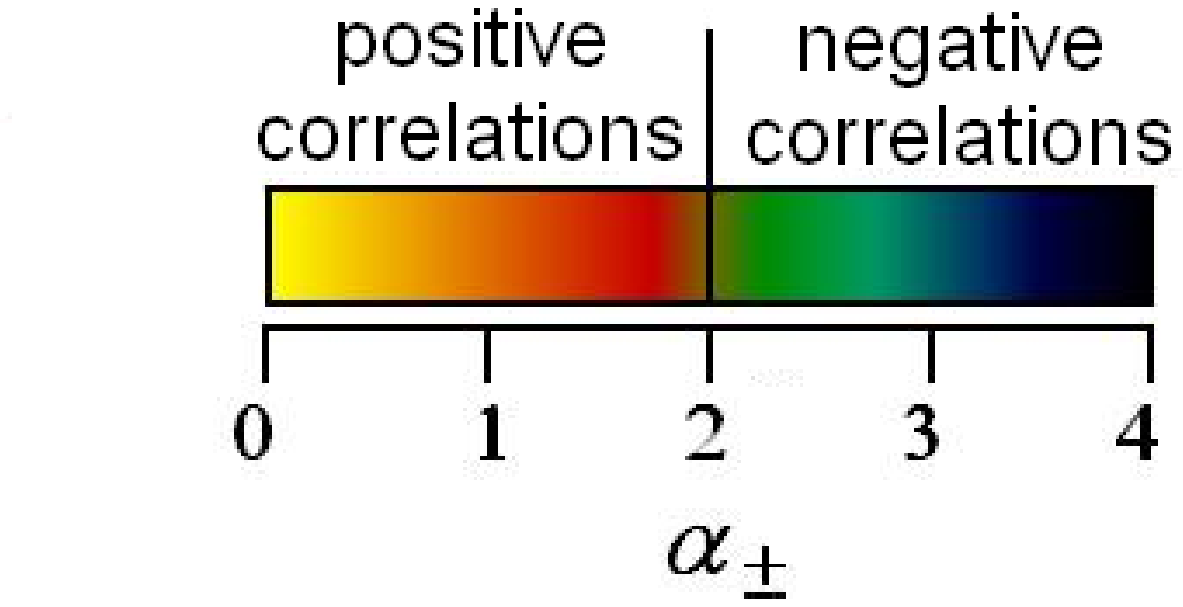} 
\caption{\label{fig7} Comparison of the correlation structure of the demise events the the quiet photosphere (a), in the NOAA 10365 active region without significant flaring activity (b) and in the same active region producing M-class flares (c). Left column: GCI exponents versus time scale $\tau$ plotted in the same format as in Fig. \ref{fig4}. Right column: continuous scalograms showing the dependence of the cross-correlation dimension $\alpha_{\pm}$ on spatial and temporal separation between bipolar demise events. The red color prevailing in the flaring active region scalogram may be a signature of the magnetic flux cancellation.}
\end{center}
\end{figure*}

In contrast, the demise events in NOAA 10365 are described by substantially lower $\alpha_{\pm}$ values signaling positive correlations in bipolar magnetic pairs, possibly due to the cancellation of flux in reconnecting magnetic loops structures. This possibility is supported by the dependence of the cross-correlation dimension on the level of coronal flaring activity as shown in Fig. \ref{fig7}. The figure presents continuous scalograms $\alpha_{\pm}(\ell, \tau)$ defined by eq. (\ref{eq5}) characterizing the magnetic flux demise dynamics in flaring and non-flaring regimes of NOAA 10365 (panels (b) and (c), correspondingly) as compared to the demise dynamics in the quiet photosphere (panel (a)). Each scalogram is accompanied by the $\tau$-dependence of small- and large-scale GCI dimensions plotted in the same format as the right column in Fig. \ref{fig4}.

The level of flaring activity is clearly encoded in the shape of the scalograms shown in Fig. \ref{fig7}. The flaring state of the active region is characterized by the condition $\alpha_{\pm} < 2$ (scalogram regions colored in red and yellow) for almost the entire studied range of $\ell$ and $\tau$, confirming the presence of positively correlated bipolar demise dynamics at various spatial and temporal scales. The sign of the correlation changes from positive to slightly negative only in two narrow bands which could be associated with finite size effects imposed by the geometry of the active region. Even for very small delay times $\tau \sim$ 1 min, there exist a distinct domain of $\ell$ scales ($\sim$ 10 to 20 Mm) where the cross-correlation is positive. As was already mentioned in Section 3.2, the quiet Sun exhibits no stable positive bipolar correlations. The narrow small-scale $\ell$ band with $\alpha_{\pm} < 2$ seen in the top right panel of Fig. \ref{fig7}. can be an artifact reflecting insufficient number of closely located bipolar pairs in the quiet-Sun network. The quiescent active region takes the intermediate position between the quiet Sun ans the flaring active region showing small but statistically significant positive correlations in selected $\ell$ ranges. As in the case with the flaring NOAA 10365, the correlations describing nearly simultaneous demise events ($\tau \approx 0$) in the non-flaring state of this active region are only detectable short spatial distances.

The observed tendencies suggest that the GCI dimensions of the active region indicative of positively correlated bipolar demise events are connected with the flaring activity. In other words, there is a coupling between the symmetric reduction of the photospheric flux and coronal dissipation. The correlated demise events could be produced by multiple bursts of magnetic reconnection in this NOAA 10365 leading to spatially localized symmetric reduction of oppositely signed line-of-sight magnetic fluxes. Indeed, the geometry of the chromospheric ribbons observed in the $H_{\alpha}$ band unambiguously shows that the flaring activity of NOAA 10365 was triggered by magnetic reconnection at coronal heights below the twisted flux tube of positive helicity \citep{chandra09}. This is consistent with the results of 3D MHD simulations of the studied active region showing the formation of multiple current sheets in the vicinity of coronal X-lines above NOAA 10365 \citep{podgorny07, podgorny08}. Each of the current sheets may well be responsible for an elementary flare. The reconnection-driven annihilation of the coronal field above NOAA 10365 could lead to a detectable reduction of the photospheric flux in magnetically connected bipolar magnetic pairs, and contribute to the observed correlated occurrence of the demise events characterized by the condition $\alpha_{\pm} <2$. To some extent, the bipolar GCI statistics can also represent simultaneous reduction of the magnetic flux in canceling bipolar pairs involved in bulk coronal heating around the active region \citep{schrij02}. 

The $\tau$ - dependence of $\alpha_{\pm}$ dimensions of the active region is different than that of the quiet Sun, with positive correlations in disappearing bipolar pairs being the strongest at smallest $\tau$ scales. This  suggests that the characteristic time scale of the apparent flux cancellation can be comparable with the 1-minute image sampling time. The correlations also increase with the decrease of the spatial scale $\ell$ and tend to be less pronounced at $\ell>\ell_0$. If the association of these events with canceling magnetic flux elements is correct, our observations indicate the dominant contribution of small- to medium size coronal loops in the reconnection process.

Positively correlated unipolar demise events are reliably identified for $B_{th}$ exceeding $\sim $2 standard deviations of the active region field but are not seen at lower detection thresholds. Therefore, they are a property of an intense magnetic field. For higher thresholds, the cross-correlations monotonously increase with $B_{th}$ suggesting an involvement of non-Gaussian magnetic field fluctuations, such as those associated with intermittent turbulence below the photosphere \citep{abramenko05} or transient multiscale dissipation in the corona \cite{uritsky07}. This explanation is in line with earlier observations showing that the preferred locations of bright coronal points correlate with the regions of intense unsigned magnetic flux (see \citet{benev10} and refs therein),
although the physical mechanism of this dependence likely to be driven by topological transitions in a complex multiscale magnetic network remains to be understood. 

\section{Conclusions}


Our findings demonstrate that the developed data analysis approach based on the GCI algorithms and direct spatiotemporal tracking of magnetic elements is capable of addressing various aspects of the dynamic of solar photospheric network, including the geometry of supergranular structures, emergence of the nonpotential flux, and cancellation dynamics of bipolar magnetic pairs. 

\begin{figure}[ht]
\noindent\includegraphics*[width=7.5 cm]{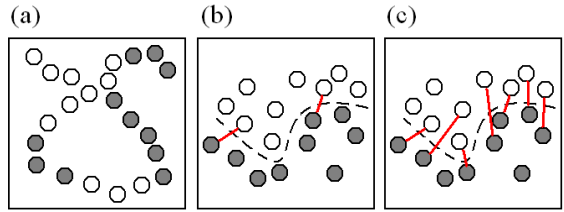} 
\caption{\label{fig8} Examples of spatial arrangement of magnetic events in the quiet Sun (a) and in the active solar region in quiescent (b) and flaring (c) states consistent with our analysis results. Red lines indicate positively correlated demise events of opposite magnetic polarity presumably associated with magnetic flux cancellation events in the active region which could be involved in sympathetic solar eruptions \citep{torok11}.}
\end{figure}

The results obtained reveal a significantly different multiscale dynamics of the photospheric magnetic flux in quiet and active solar regions as represented by the studied sample of SOHO MDI images. Quiet Sun exhibits robust unipolar clustering of both OE and DE events over a broad range of photospheric distances corresponding to meso- and supergranulation, with a small-scale quasi one-dimensional clustering along supergranular boundaries and a more isotropic clustering at the level of supergranules. These clustering regimes place important constraints on the topology of the photospheric network and spatially distributed storage of free magnetic energy in the quiet Sun as illustrated by Fig. \ref{fig8}(a). No detectable cross-correlations in bipolar magnetic pairs which could be related with unstable magnetic loops were found in the quiet photospheric region. NOAA 10365 solar active region exhibits signatures of non-potential magnetic structures with a topologically complex polarity separation line. 

Flaring activity in the active region is accompanied by positively correlated magnetic demise events which can be a signature of magnetic reconnection. A possible underlying geometry which would lead to the observed CGI signatures of the active region is a sheared loop arcade with an irregular polarity inversion line as depicted in Fig. \ref{fig8}(b,c). The topological complexity of this structure can be instrumental because simple shearing of an arcade is typically insufficient to any significant manifestation of solar activity such as flares or prominences \citep{dahlburg91}. The statistical results obtained for AR 103675 speak in favor of so-called sympathetic eruptions in this complex active region. Such eruptions may occur within a relatively short period of time in spatially separated flux ropes due to a variety of MHD-scale triggers destabilizing the coronal field \citep{torok11}. 

The proposed methodology can be helpful in resolving a broader set of theoretical and practical problems associated with solar magnetic complexity and its impact on solar activity. For example, is there a relationship between statistical properties of fractality/complexity in the photosphere and corona? Some studies provide definitive answer (e.g., \citet{conlon10, falconer09, abramenko10, abramenko10a}), but some others argues against such possibility \citep{dimitropoulou09}. If the complexity linking between the two solar structures exists, one can expect the CGI signatures of photosphere and corona to vary in a consistent way. 

It is also not clear how the fundamental plasma processes responsible for the formation and evolution of the magnetic network organize themselves across such a broad range of scales. Accurate evaluation of CGI dimensions for statistically representative sets of solar regions and conditions may provide important clues on the physics of the underlying process, for instance, by validating certain types of turbulent cascades \citep{muller00}. The proposed methodology can also help to better understand how the scaling properties of the quiet and active Sun networks relate to each other, and how these properties reflect the formation of unstable magnetic topologies leading to plasma heating and dissipation. The relationship between the temporal and spatial scaling laws governing multiscale geometry and dynamics of solar magnetic field is another important subject of future research. Scaling characteristics of sheared magnetic structures generated by oppositely directed line-of-sight flux elements, and the contribution of these structures the coronal dissipation in quiet and active solar regions are also largely unexplored and requires careful consideration. To clarify the underlying physics of the observed effects, it would be essential to combine our statistical tools with conjugate observations of coronal activity as well as solar tomography \citep{davila94} and other three-dimensional reconstruction techniques. 

The developed approach can also be used for improving the forecasting of flaring activity through identification of topologically complex unstable magnetic configurations. Spatial magnetic complexity of an active region as measured by the power-law decay of the Fourier spectrum of the longitudinal field was shown to correlate with the flare productivity of the region \citep{abramenko05}. This dependence suggests that active regions with more intermittent magnetic structure tend to produce higher energy flaring events. Our methodology offers a more comprehensive way of measuring magnetic intermittency characterizing  each magnetic polarity (based on $\alpha_+$ and $\alpha_-$), as well as bipolar magnetic structures ($\alpha_{\pm}$). Analysis of these correlation measures, as well as the temporal CGI dimensions not considered here, may help better understand photospheric drivers of bulk coronal heating and local topological precursors of bright coronal points and flaring activity across the entire range of the involved spatial and temporal scales. 

\acknowledgments

We want to thank V. Abramenko for preparing the set of SOHO MDI high resolution magnetograms used in this study and valuable comments on the manuscript. We are also grateful to S. Antiochos, D. Falconer, J. Gurman, K. Muglach, D. Pesnell, D. Rabin, V. Titov and V. Yurchyshyn for helpful comments and advice.  This work was partly supported by the contract WAP 612.1-003 from ADNET Systems. The code of the stochastic Cantor dust set with adjustable fractal dimension was developed by G. Uritskiy.





\begin{thebibliography}{80}
\expandafter\ifx\csname natexlab\endcsname\relax\def\natexlab#1{#1}\fi

\bibitem[{Abramenko(2005)}]{abramenko05}
Abramenko, V.~I. 2005, Astrophys. J., 629, 1141

\bibitem[{Abramenko \& Longcope(2005)}]{abramenko05a}
Abramenko, V.~I., \& Longcope, D.~W. 2005, Astrophysical J., 619, 1160

\bibitem[{Abramenko \& Yurchyshyn(2010{\natexlab{a}})}]{abramenko10}
Abramenko, V.~I., \& Yurchyshyn, V.~B. 2010{\natexlab{a}}, Astrophysical J.,
  722, 122

\bibitem[{Abramenko \& Yurchyshyn(2010{\natexlab{b}})}]{abramenko10a}
---. 2010{\natexlab{b}}, Astrophysical J., 720, 717

\bibitem[{Abramenko {et~al.}(2002)Abramenko, Yurchyshyn, Wang, Spirock, \&
  Goode}]{abramenko02}
Abramenko, V.~I., Yurchyshyn, V.~B., Wang, H., Spirock, T.~J., \& Goode, P.~R.
  2002, Astrophysical J., 577, 487

\bibitem[{Ahmed {et~al.}(2010)Ahmed, Qahwaji, Colak, Wit, \& Ipson}]{ahmed10}
Ahmed, O.~W., Qahwaji, R., Colak, T., Wit, T. D.~D., \& Ipson, S. 2010, Visual
  Computer, 26, 385

\bibitem[{Antiochos(1998)}]{antiochos98}
Antiochos, S.~K. 1998, Astrophysical J., 502, L181

\bibitem[{Antiochos {et~al.}(2007)Antiochos, DeVore, Karpen, \&
  Mikic}]{antiochos07}
Antiochos, S.~K., DeVore, C.~R., Karpen, J.~T., \& Mikic, Z. 2007,
  Astrophysical J., 671, 936

\bibitem[{Aschwanden(2006)}]{aschwanden06}
Aschwanden, M.~J., ed. 2006, Physics of the solar corona (Berlin Heidelberg:
  Springer-Verlag)

\bibitem[{Aschwanden(2011)}]{aschwanden11}
---. 2011, Self-organized criticality in astrophysics (Berlin Heidelberg:
  Springer-Verlag)

\bibitem[{Aschwanden \& Parnell(2002)}]{aschwanden02}
Aschwanden, M.~J., \& Parnell, C.~E. 2002, Astrophysical J., 572, 1048

\bibitem[{Balke {et~al.}(1993)Balke, Schrijver, Zwaan, \& Tarbell}]{balke93}
Balke, A.~C., Schrijver, C.~J., Zwaan, C., \& Tarbell, T.~D. 1993, Solar Phys.,
  143, 215

\bibitem[{Barabasi \& Stanley(1995)}]{barabasi95}
Barabasi, A.~L., \& Stanley, H.~E., eds. 1995 (Cambridge, UK: Cambridge
  University Press)

\bibitem[{Benevolenskaya(2010)}]{benev10}
Benevolenskaya, E.~E. 2010, Astronomische Nachrichten, 331, 63

\bibitem[{Berger {et~al.}(1998)Berger, Lofdahl, Shine, \& Title}]{berger98}
Berger, T.~E., Lofdahl, M.~G., Shine, R.~S., \& Title, A.~M. 1998,
  Astrophysical J., 495, 973

\bibitem[{Berger \& Title(1996)}]{berger96}
Berger, T.~E., \& Title, A.~M. 1996, Astrophysical J., 463, 365

\bibitem[{Cadavid {et~al.}(1994)Cadavid, Lawrence, Ruzmaikin, \&
  Kaylengknight}]{cadavid94}
Cadavid, A.~C., Lawrence, J.~K., Ruzmaikin, A.~A., \& Kaylengknight, A. 1994,
  Astrophysical J., 429, 391

\bibitem[{Chae {et~al.}(2004)Chae, Moon, \& Park}]{chae04}
Chae, J., Moon, Y.~J., \& Park, Y.~D. 2004, Solar Phys., 223, 39

\bibitem[{Chandra {et~al.}(2009)Chandra, Schmieder, Aulanier, \&
  Malherbe}]{chandra09}
Chandra, R., Schmieder, B., Aulanier, G., \& Malherbe, J.~M. 2009, Solar Phys.,
  258, 53

\bibitem[{Chaouche {et~al.}(2011)Chaouche, Moreno-Insertis, Pillet, Wiegelmann,
  Bonet, Knolker, Rubio, Iniesta, Barthol, Gandorfer, Schmidt, \&
  Solanki}]{chaouche11}
Chaouche, L.~Y., {et~al.} 2011, Astrophysical J. Lett., 727, L30

\bibitem[{Charbonneau {et~al.}(2001)Charbonneau, McIntosh, Liu, \&
  Bogdan}]{charbonneau01}
Charbonneau, P., McIntosh, S.~W., Liu, H.~L., \& Bogdan, T.~J. 2001, Solar
  Phys., 203, 321

\bibitem[{Conlon {et~al.}(2010)Conlon, McAteer, Gallagher, \&
  Fennell}]{conlon10}
Conlon, P.~A., McAteer, R. T.~J., Gallagher, P.~T., \& Fennell, L. 2010,
  Astrophysical J., 722, 577

\bibitem[{Crouch {et~al.}(2007)Crouch, Charbonneau, \& Thibault}]{crouch07}
Crouch, A.~D., Charbonneau, P., \& Thibault, K. 2007, Astrophysical J., 662,
  715

\bibitem[{Dahlburg {et~al.}(1991)Dahlburg, Antiochos, \& Zang}]{dahlburg91}
Dahlburg, R.~B., Antiochos, S.~K., \& Zang, T.~A. 1991, Astrophysical J., 383,
  420

\bibitem[{Davila(1994)}]{davila94}
Davila, J.~M. 1994, Astrophysical J., 423, 871

\bibitem[{DeForest {et~al.}(2007)DeForest, Hagenaar, Lamb, Parnell, \&
  Welsch}]{deforest07}
DeForest, C.~E., Hagenaar, H.~J., Lamb, D.~A., Parnell, C.~E., \& Welsch, B.~T.
  2007, Astrophys. J., 666, 576

\bibitem[{Dimitropoulou {et~al.}(2009)Dimitropoulou, Georgoulis, Isliker,
  Vlahos, Anastasiadis, Strintzi, \& Moussas}]{dimitropoulou09}
Dimitropoulou, M., Georgoulis, M., Isliker, H., Vlahos, L., Anastasiadis, A.,
  Strintzi, D., \& Moussas, X. 2009, Astronomy and Astrophysics, 505, 1245

\bibitem[{Edmondson {et~al.}(2010)Edmondson, Antiochos, DeVore, \&
  Zurbuchen}]{edmondson10}
Edmondson, J.~K., Antiochos, S.~K., DeVore, C.~R., \& Zurbuchen, T.~H. 2010,
  Astrophysical J., 718, 72

\bibitem[{Falconer {et~al.}(2008)Falconer, Moore, \& Gary}]{falconer08}
Falconer, D.~A., Moore, R.~L., \& Gary, G.~A. 2008, Astrophysical J., 689, 1433

\bibitem[{Falconer {et~al.}(2009)Falconer, Moore, Gary, \& Adams}]{falconer09}
Falconer, D.~A., Moore, R.~L., Gary, G.~A., \& Adams, M. 2009, Astrophysical J.
  Lett., 700, L166

\bibitem[{Falconer {et~al.}(1997)Falconer, Moore, Porter, Gary, \&
  Shimizu}]{falconer97}
Falconer, D.~A., Moore, R.~L., Porter, J.~G., Gary, G.~A., \& Shimizu, T. 1997,
  Astrophysical J., 482, 519

\bibitem[{Falconer {et~al.}(1998)Falconer, Moore, Porter, \&
  Hathaway}]{falconer98}
Falconer, D.~A., Moore, R.~L., Porter, J.~G., \& Hathaway, D.~H. 1998,
  Astrophysical J., 501, 386

\bibitem[{Georgoulis(2005)}]{georgoulis05}
Georgoulis, M.~K. 2005, Solar Phys., 228, 5

\bibitem[{Grassberger(1983)}]{grassberger83a}
Grassberger, P. 1983, Phys. Lett. A, 97, 227

\bibitem[{Grassberger \& Procaccia(1983)}]{grassberger83}
Grassberger, P., \& Procaccia, I. 1983, Physica D, 9, 189

\bibitem[{Hagenaar {et~al.}(1997)Hagenaar, Schrijver, \& Title}]{hagenaar97}
Hagenaar, H.~J., Schrijver, C.~J., \& Title, A.~M. 1997, Astrophysical J., 481,
  988

\bibitem[{Hathaway {et~al.}(2000)Hathaway, Beck, Bogart, Bachmann, Khatri,
  Petitto, Han, \& Raymond}]{hathaway00}
Hathaway, D.~H., Beck, J.~G., Bogart, R.~S., Bachmann, K.~T., Khatri, G.,
  Petitto, J.~M., Han, S., \& Raymond, J. 2000, Solar Phys., 193, 299

\bibitem[{Hathaway {et~al.}(2002)Hathaway, Beck, Han, \& Raymond}]{hathaway02}
Hathaway, D.~H., Beck, J.~G., Han, S., \& Raymond, J. 2002, Solar Phys., 205,
  25

\bibitem[{Hughes {et~al.}(2003)Hughes, Paczuski, Dendy, Helander, \&
  {McClements}}]{hughes03}
Hughes, D., Paczuski, M., Dendy, R.~O., Helander, P., \& {McClements}, K.~G.
  2003, Phys. Rev. Lett., 90, 131101

\bibitem[{Kantz(1994)}]{kantz94}
Kantz, H. 1994, Phys. Rev. E, 49, 5091

\bibitem[{Klimchuk(2006)}]{klimchuk06}
Klimchuk, J.~A. 2006, Solar Phys., 234, 41

\bibitem[{Lamb {et~al.}(2008)Lamb, DeForest, Hagenaar, Parnell, \&
  Welsch}]{lamb08}
Lamb, D.~A., DeForest, C.~E., Hagenaar, H.~J., Parnell, C.~E., \& Welsch, B.~T.
  2008, Astrophys. J., 674, 520

\bibitem[{Lawrence {et~al.}(1993)Lawrence, Ruzmaikin, \& Cadavid}]{lawrence93}
Lawrence, J.~K., Ruzmaikin, A.~A., \& Cadavid, A.~C. 1993, Astrophysical J.,
  417, 805

\bibitem[{Leighton {et~al.}(1962)Leighton, Noyes, \& Simon}]{leighton62}
Leighton, R.~B., Noyes, R.~W., \& Simon, G.~W. 1962, Astrophysical J., 135, 474

\bibitem[{Lisle {et~al.}(2000)Lisle, Rosa, \& Toomre}]{lisle00}
Lisle, J., Rosa, M.~D., \& Toomre, J. 2000, Solar Phys., 197, 21

\bibitem[{Lu \& Hamilton(1991)}]{lu91}
Lu, E.~T., \& Hamilton, R.~J. 1991, Astrophysical J., 380, L89

\bibitem[{Martin(1984)}]{martin84}
Martin, S.~F. 1984, in Small-scale dynamical processes in quiet stellar
  atmospheres, ed. S.~L. Keil, National Solar Observatory, Sunspot, NM, 30

\bibitem[{Martin(1988)}]{martin88}
Martin, S.~F. 1988, Solar Phys., 117, 243

\bibitem[{McAteer {et~al.}(2010)McAteer, Gallagher, \& Conlon}]{mcateer10}
McAteer, R. T.~J., Gallagher, P.~T., \& Conlon, P.~A. 2010, Adv. Space Res.,
  45, 1067

\bibitem[{Meunier {et~al.}(2010)Meunier, Lagrange, \& Desort}]{meunier10}
Meunier, N., Lagrange, A.~M., \& Desort, M. 2010, Astronomy and Astrophysics,
  519, A66

\bibitem[{Morales \& Charbonneau(2008)}]{morales08}
Morales, L., \& Charbonneau, P. 2008, Astrophysical J., 682, 654

\bibitem[{Muller \& Biskamp(2000)}]{muller00}
Muller, W.~C., \& Biskamp, D. 2000, Phys. Rev. Lett., 84, 475

\bibitem[{Parnell {et~al.}(2009)Parnell, DeForest, Hagenaar, Johnston, Lamb, \&
  Welsch}]{parnell09}
Parnell, C.~E., DeForest, C.~E., Hagenaar, H.~J., Johnston, B.~A., Lamb, D.~A.,
  \& Welsch, B.~T. 2009, Astrophysical J., 698, 75

\bibitem[{Podgorny \& Podgorny(2008)}]{podgorny08}
Podgorny, A.~I., \& Podgorny, I.~M. 2008, Astronomy Reports, 52, 666

\bibitem[{Podgorny {et~al.}(2007)Podgorny, Podgorny, \&
  Meshalkina}]{podgorny07}
Podgorny, A.~I., Podgorny, I.~M., \& Meshalkina, N.~S. 2007, Solar System
  Research, 41, 322

\bibitem[{Rast(2003)}]{rast03}
Rast, M.~P. 2003, Astrophysical J., 597, 1200

\bibitem[{Scherrer {et~al.}(1995)Scherrer, Bogart, Bush, Hoeksema, Kosovichev,
  Schou, Rosenberg, Springer, Tarbell, Title, Wolfson, Zayer, Akin, Carvalho,
  Chevalier, Duncan, Edwards, Katz, Levay, Lindgren, Mathur, Morrison, Pope,
  Rehse, \& Torgerson}]{scherrer95}
Scherrer, P.~H., {et~al.} 1995, Solar Phys., 162, 129

\bibitem[{Schrijver {et~al.}(2005)Schrijver, DeRosa, Title, \&
  Metcalf}]{schrij05}
Schrijver, C.~J., DeRosa, M.~L., Title, A.~M., \& Metcalf, T.~R. 2005,
  Astrophys. J., 628, 501–

\bibitem[{Schrijver \& Title(2002)}]{schrij02}
Schrijver, C.~J., \& Title, A.~M. 2002, Solar Phys., 207, 223–

\bibitem[{Schrijver {et~al.}(1997)Schrijver, Title, Ballegooijen, Hagenaar, \&
  Shine}]{schrij97}
Schrijver, C.~J., Title, A.~M., Ballegooijen, A. A.~V., Hagenaar, H.~J., \&
  Shine, R.~A. 1997, Astrophys. J., 487, 424

\bibitem[{Schrijver {et~al.}(1992)Schrijver, Zwaan, Balke, Tarbell, \&
  Lawrence}]{schrijver92}
Schrijver, C.~J., Zwaan, C., Balke, A.~C., Tarbell, T.~D., \& Lawrence, J.~K.
  1992, Astronomy and Astrophysics, 253, L1

\bibitem[{Schuster \& Wolfram(2005)}]{schuster05}
Schuster, H.~G., \& Wolfram, J., eds. 2005, Deterministic chaos: an
  introduction (Weinheim: Wiley-VCH Verlag)

\bibitem[{Settele {et~al.}(2002)Settele, Carroll, Nickelt, \&
  Norton}]{settele02}
Settele, A., Carroll, T.~A., Nickelt, I., \& Norton, A.~A. 2002, Astronomy and
  Astrophysics, 386, 1123

\bibitem[{Shine {et~al.}(2000)Shine, Simon, \& Hurlburt}]{shine00}
Shine, R.~A., Simon, G.~W., \& Hurlburt, N.~E. 2000, Solar Phys., 193, 313

\bibitem[{Simon {et~al.}(1995)Simon, Title, \& Weiss}]{simon95}
Simon, G.~W., Title, A.~M., \& Weiss, N.~O. 1995, Astrophysical J., 442, 886

\bibitem[{Simon {et~al.}(2001)Simon, Title, \& Weiss}]{simon01}
---. 2001, Astrophysical J., 561, 427

\bibitem[{Solanki {et~al.}(2006)Solanki, Inhester, \& Schussler}]{solanki06}
Solanki, S.~K., Inhester, B., \& Schussler, M. 2006, Reports on Progress in
  Phys., 69, 563

\bibitem[{Srikanth {et~al.}(2000)Srikanth, Singh, \& Raju}]{srikanth00}
Srikanth, R., Singh, J., \& Raju, K.~P. 2000, Astrophysical J., 534, 1008

\bibitem[{Tian {et~al.}(2011)Tian, Demoulin, Alexander, \& Zhu}]{tian11}
Tian, L.~R., Demoulin, P., Alexander, D., \& Zhu, C.~M. 2011, Astrophysical J.,
  727, 28

\bibitem[{T\"{o}r\"{o}k {et~al.}(2011)T\"{o}r\"{o}k, Panasenco, Titov, Mikic,
  Reeves, Velli, Linker, \& Toma}]{torok11}
T\"{o}r\"{o}k, T., Panasenco, O., Titov, V.~S., Mikic, Z., Reeves, K.~K.,
  Velli, M., Linker, J.~A., \& Toma, G.~D. 2011, Astrophysical J. Lett., 739,
  L63

\bibitem[{Turcotte(1989)}]{turcotte89}
Turcotte, D.~L. 1989, Pure Appl. Geophysics, 131, 171

\bibitem[{Uritsky {et~al.}(2010{\natexlab{a}})Uritsky, Donovan, Trondsen,
  Pineau, \& Kozelov}]{uritsky10}
Uritsky, V.~M., Donovan, E., Trondsen, T., Pineau, D., \& Kozelov, B.~V.
  2010{\natexlab{a}}, J. Geophysical Research - Space Phys., 115, A09205

\bibitem[{Uritsky {et~al.}(2007)Uritsky, Paczuski, Davila, \&
  Jones}]{uritsky07}
Uritsky, V.~M., Paczuski, M., Davila, J.~M., \& Jones, S.~I. 2007, Phys. Rev.
  Lett., 99, 025001

\bibitem[{Uritsky {et~al.}(2010{\natexlab{b}})Uritsky, Pouquet, Rosenberg,
  Mininni, \& Donovan}]{uritsky10a}
Uritsky, V.~M., Pouquet, A., Rosenberg, D., Mininni, P.~D., \& Donovan, E.~F.
  2010{\natexlab{b}}, Phys. Rev. E, 82, 056326

\bibitem[{Uritsky {et~al.}(2011)Uritsky, Slavin, Khazanov, Donovan, Boardsen,
  Anderson, \& Korth}]{uritsky11}
Uritsky, V.~M., Slavin, J.~A., Khazanov, G.~V., Donovan, E.~F., Boardsen,
  S.~A., Anderson, B.~J., \& Korth, H. 2011, J. Geophysical Research -- Space
  Phys., 116, A09236

\bibitem[{Verbeeck {et~al.}(2011)Verbeeck, Higgins, Colak, Watson, Delouille,
  Mampaey, \& Qahwaji}]{verbeeck11}
Verbeeck, C., Higgins, P.~A., Colak, T., Watson, F.~T., Delouille, V., Mampaey,
  B., \& Qahwaji, R. 2011, Solar Physics (submitted), arXiv:1109.0473v1

\bibitem[{Viall \& Klimchuk(2011)}]{viall11}
Viall, N.~M., \& Klimchuk, J.~A. 2011, Astrophysical J., 738, 24

\bibitem[{Wang {et~al.}(1996)Wang, Tang, Zirin, \& Wang}]{wang96}
Wang, H., Tang, F., Zirin, H., \& Wang, J. 1996, Solar Phys., 165, 223

\bibitem[{Welsch {et~al.}(2012)Welsch, Kusano, Yamamoto, \& Muglach}]{welsch12}
Welsch, B.~T., Kusano, K., Yamamoto, T.~T., \& Muglach, K. 2012, Astrophys. J.,
  in press

\bibitem[{Williams \& Pesnell(2011)}]{williams11}
Williams, P.~E., \& Pesnell, W.~D. 2011, Solar Phys., 270, 125

\end{thebibliography}




\end{document}